\documentclass[12pt,preprint]{aastex}

\shorttitle{Low Mass White Dwarfs}
\shortauthors{Kilic, Allende Prieto, \& Brown}

\begin{document}

\title{The Lowest Mass White Dwarf}

\author{Mukremin Kilic\altaffilmark{1}, Carlos Allende Prieto\altaffilmark{2}, Warren R. Brown\altaffilmark{3}, and D. Koester\altaffilmark{4}}

\altaffiltext{1}{Columbus Fellow, Department of Astronomy, Ohio State University, 140 West 18th Avenue, Columbus, OH 43210, USA; kilic@astronomy.ohio-state.edu}

\altaffiltext{2}{McDonald Observatory and Department of Astronomy, University of Texas, Austin, TX 78712, USA; callende@astro.as.utexas.edu}

\altaffiltext{3}{Smithsonian Astrophysical Observatory, 60 Garden Street, Cambridge, MA 02138, USA; wbrown@cfa.harvard.edu}

\altaffiltext{4}{Institut f\"ur Theoretische Physik und Astrophysik, University of Kiel, 24098 Kiel, Germany; koester@astrophysik.uni-kiel.de}

\begin{abstract}

Extremely low mass white dwarfs are very rare objects likely formed in compact binary systems.
We present MMT optical spectroscopy of 42 low mass white dwarf
candidates serendipitously discovered in a survey for hypervelocity B-type
stars. One of these objects, SDSS J0917+46, has $T_{\rm eff}=$ 11,288 $\pm$ 72 K and $\log$ g = 5.48 $\pm$ 0.03;
with an estimated mass of 0.17 $M_\sun$, it is the lowest gravity/mass
white dwarf currently known. However, 40 of the low mass candidates are normal DA white
dwarfs with apparently inaccurate SDSS $g$ magnitudes. We revisit the
identification of low mass white dwarf candidates previously found in the
SDSS, and conclude that four objects have $M < 0.2 M_\sun$. None of these
white dwarfs show excess emission from a binary companion, and radial
velocity searches will be necessary to constrain the nature of the unseen
companions.

\end{abstract}


\section{Introduction}

The mass distribution of hot DA (hydrogen-rich atmosphere) white dwarfs (WDs) from the Palomar Green Survey peaks at 0.57 $M_\sun$ with
a dispersion of 0.19 $M_\sun$ (Liebert et al. 2005). A similar mass distribution is also observed for WDs cooler than 12,000 K
(Bergeron et al. 2001). In both cases, separate low mass and high mass components to this mass distribution are also observed.
The Liebert et al. sample includes WDs with masses as low as 0.32 $M_\sun$, and the mass distribution has a low mass peak
at 0.40 $M_\sun$. These low mass WDs are understood as He-core WDs formed in binary systems. 
He-core WDs can be formed when a companion strips the outer envelope from a post main sequence star
before the star reaches the tip of the red giant branch and ignites the helium. Low mass WDs are usually found
in close binaries, mostly double degenerate systems (Marsh et al. 1995).
The Galaxy is not old enough to produce these WDs through single-star evolution.

Recent discoveries of extremely low mass ($\log$ g $<$ 7, hereafter ELM) WDs in the field and around pulsars show that some of these
WDs retain only a small fraction of their progenitor mass and end up with as little mass as 0.2 $M_\sun$. Liebert et al.
(2004) were the first ones to discover an ELM WD in the field (SDSS J1234-02) with $T_{\rm eff}=$
17,470 $\pm$ 750 K and $\log$ g = 6.38 $\pm$ 0.05. Eisenstein et al. (2006) increased the number of ELM WD candidates found
in the Sloan Digital Sky Survey (SDSS) to 13, including two objects with $\log$ g $<$ 6. Kawka et al. (2006) added one more object to the list (LP400-22) with
$\log$ g = 6.32 $\pm$ 0.08. The optical photometry of none of these objects shows an excess, and the nature of the companions to these
objects is not known.

The existence of ELM WDs around pulsars suggests that neutron star companions may be responsible for
creating such low mass WDs. PSR J0437-4715, J0751+1807, J1012+5307, J1713+0747, B1855+09, and J1909-3744 are pulsars in
circular pulsar-He WD binary systems with orbital periods of $\sim$0.2-100 days (Nice et al. 2005).
The properties of the pulsars in these systems differ markedly from those of typical isolated pulsars, showing more rapid spin periods
and smaller inferred magnetic fields. They also tend to have masses greater than the canonical value of 1.35 $M_\sun$. The relatively
high masses of pulsars in these systems presumably result from extended mass accretion during the late stages of their evolution (Nice
et al. 2005). At the cessation of mass transfer, the neutron star turns on as a millisecond radio pulsar and the companion is left
as a helium-core WD (van Kerkwijk et al. 2000).

Optical spectroscopy of two of the low mass WD companions to pulsars are available in the literature. van Kerkwijk et al. (1996) found a
best fit solution of $T_{\rm eff}=$ 8,550 $\pm$ 25 K and $\log$ g = 6.75 $\pm$ 0.07 for the companion to PSR J1012+5307, whereas
Bassa et al. (2006) measured $T_{\rm eff}=$ 10,090 $\pm$ 150 K and $\log$ g = 6.44 $\pm$ 0.20 for the companion to PSR J1911-5958A. 
In addition, Heber et al. (2003) found a subluminous B (sdB) star, HD 188112, with $T_{\rm eff}=$ 21500 $\pm$ 500 K and $\log$ g =
5.66 $\pm$ 0.05 and suggested that it is the progenitor of a helium core WD. 
HD 188112 is in a binary system with an $M > 0.73M_\sun$ compact object, a WD or a neutron star.

Recently, Brown et al. (2006) performed a search for hypervelocity stars in the SDSS Fourth Data Release. They used
the SDSS photometry to select faint B star candidates in a survey area of $\sim$3000 deg$^2$. 
They obtained follow-up spectroscopy of 247 candidates with the Blue Channel Spectrograph on the 6.5m MMT telescope, 44 of which
turned out to be WDs. The SDSS colors for ELM WDs fall into the same region as their selection
region for B type stars, thus they identify all of these WDs as potential low mass WDs with $\log$ g $<$ 7.
Figure 1 shows $u-g$ versus $g-r$ color-color diagram of the low mass WD candidates from Brown et al. (2006; circles)
and Eisenstein et al. (2006; triangles). Our synthetic photometry of WD model atmospheres with $\log$ g = 5, 6, 7, and 8
(Finley et al. 1997) are shown as solid lines. B type stars found in Brown et al.'s survey are shown as crosses.
It is clear from this figure that all of the WDs found by Brown et al. (2006) lie to the right of the $\log$ g = 7 line and
are candidate low mass WDs. In addition, one of these WDs has colors similar to the lowest mass WD candidate identified by Eisenstein
et al. (2006).

There are currently 5 confirmed ELM WDs (SDSS J1234-02, LP400-22, HD 188112, and companions to PSR J1012+5307 and
J1911-5958A), and 12 more SDSS candidates identified by Eisenstein et al. (2006). In this paper, we analyze the spectra
of the low mass WD candidates observed at the MMT by performing temperature and surface gravity fits to grids of pure hydrogen
atmosphere models of Finley et al. (1997). We also revisit the model fits to the Eisenstein et al. (2006) low mass WD sample.
Our fitting procedures and the measured parameters for the MMT sample are discussed
in \S 2. \S 3 describes our efforts to understand the SDSS low mass WD sample, and results from this analysis
are discussed in \S 4. 

\section{The MMT Sample}

Brown et al. (2006) identified 44 WDs, including 42 DA WDs, a helium-rich (DB) WD, and another metal-rich (DZ) WD with
Ca H and K lines. Here we exclude the two non-DA WDs from our analysis. The SDSS positions and de-reddened photometry for our sample 
are given in Table 1. The $g$-band magnitudes for these WDs range from 17.3 to 20.1. The exposure times for the
MMT spectroscopy were in the range 210 -- 1800 seconds. Most spectra were obtained with a 1.25$\arcsec$ slit yielding a resolving power of $R=$ 3500,
however several of them were obtained with a 1.0$\arcsec$ or 1.5$\arcsec$ slit that resulted in $R=$ 4300 or 2900. 
All spectra were obtained at the parallactic angle. The spectra were flux-calibrated using the blue
spectrophotometric standards (Massey et al. 1988).
Repeat observations of one of our WD candidates over 4 nights showed that the wavelength dependent variations in the continuum shape
are on the order of 6\% for our typical WD spectra. Hence, the relative flux levels at different
wavelengths are reliable at the 6\% level.  

\subsection{Spectroscopic Analysis}

The first step in our analysis is to smooth the observed spectra to a common resolving power in order to compare them with the same set of
WD model atmospheres. We have chosen to smooth the spectra to $R=$ 2500. This also
increases slightly the signal-to-noise ratio ($S/N$) of the observed spectra. We use the radial velocities measured by Brown et al. (2006)
to correct the spectra to the rest frame. After smoothing the spectra by convolution with a Gaussian kernel of the appropriate width, and resampling
with 4 points per resolution element (constant
step in $\log \lambda$), the spectra were normalized using two different procedures: $a)$ determining the continuum by iteratively
fitting a polynomial (6 iterations), then
dividing by the fit, and $b)$ dividing each spectrum by its median value. In the first case, the shape of the spectral energy
distribution (SED) is effectively removed, losing valuable information but avoiding systematic errors in the flux calibration and the effects of
reddening. We note that we apply the same procedures to both the observed and model spectra in all cases.

The grid of WD model atmospheres covers effective temperatures from 6,000 K to 30,000 K in steps
of 1,000 K, and surface gravities from $\log$ g = 5.0 to 9.0 in steps of 0.25 dex. The best-fit parameters are determined for each
star by searching for the minimum $\chi^2$ in the range covered by the spectrum using the Nelder-Mead algorithm (Nelder \& Mead 1965), performing quadratic
Bezier interpolation of the spectra in both $T_{\rm eff}$ and $\log$ g (Allende Prieto 2004). In the case of full continuum normalization (case $a$), we only fit 
the Balmer line profiles, whereas we fit the entire spectrum in case $b$. The derived error bars for the parameters are simply the square roots of the 
diagonal elements of the covariance matrix (Press et al. 1986).
 
If the analysis is carried out at different resolving powers, the results change slightly for case $a$ where we fit the Balmer lines. We have
repeated our fits with $R$ changing from 2500 to 200, and found that the drift can be approximately matched with a 1/$R$ law. We find that
a reduction of resolving power from infinity to 2500 causes $T_{\rm eff}$ to be over-estimated by 1\% and $\log$ g by 0.04 dex. 
Since our spectra are mostly composed of strong Balmer lines, it is hard to determine the continuum in the blue part of the
spectrum. In addition, the removal of the continuum sacrifices useful information and, in the presence of noise, introduces
degeneracy -- a decrease in $T_{\rm eff}$ can be compensated by an increase in $\log$ g, and multiple minima are apparent, although
one is usually significantly deeper than the other.  

In case $b$ where we fit the entire spectrum, the parameters that we derive are independent of the resolution that we use down to
$R>$ 200, therefore these results are more robust than those derived from the normalized spectra. In addition, the degeneracy problem is
resolved with the additional information contained in the continuum shape. A comparison of our fits for cases $a$ and $b$ shows that
on average our temperature and $\log$ g estimates differ by $\sim$800 K and 0.07 dex, respectively.
Figure 2 shows the $\log$ g for our spectral fits using the line profiles from the continuum corrected
spectra (case $a$) versus using the spectra without any continuum correction (case $b$). This figure suggests that our gravity estimates
are robust. With the exception of four objects for which flux calibration errors are apparent in the observed spectra (Cand017, 020, 124, and 192\footnote{Cand192
has an apparent Ca I line at 4226 \AA. Ca II lines may be present in the spectrum, but not easily recognizable due to noisier spectrum in
the blue. Our best fit $T_{\rm eff}=$ 13,000 K solution suggests that if the calcium was photospheric, we would expect to see strong Ca II lines in the
spectrum of this object (see the optical spectrum of the metal rich WD GD362 in Gianninas et al. 2004). The observed Ca I line may come from a late type
companion to this otherwise normal WD.}),
the quality of our fits (minimum $\chi^2$) is always better when the continuum shape is not removed.

We prefer the $T_{\rm eff}$ and $\log$ g from our best fit case $b$ solutions.
Our fits to the original spectra contain more information than
the continuum corrected spectra which help us resolve the $T_{\rm eff}$ and $\log$ g degeneracy problem, and
the solutions are independent of the spectral resolution used. Our choice of spectral normalization (case $b$
over case $a$) does not change our results significantly (see Figure 2).
Table 1 presents our adopted values of $T_{\rm eff}$, $\log$ g, and $\chi^2$. Figure 3 shows our best-fit model spectra (red lines) to the observed
spectra (black lines) of low mass WD candidates identified by Brown et al. (2006). 

Repeat observations of one of our targets over several nights shows that
our real errors are larger than our internal error estimates (in Table 1).
Analyzing 9 different spectra of the same object,
we find that the average internal errors in $T_{\rm eff}$ and $\log$ g are
40 K and 0.02 dex, respectively, while the empirical estimates indicate 600 K and 0.08 dex. 
The real error bars (set by systematic errors in our analysis) are more like 5\% in $T_{\rm eff}$ and 0.08 dex in $\log$ g.

\subsection{Results}

A glance at our best-fit $T_{\rm eff}$ and $\log$ g values given in Table 1 shows that the majority of the objects in our sample have $\log$
g $>$ 7 and therefore, by our definition, are not ELM WDs. This is completely at odds with the SDSS photometry, and we will come back to this point in the next
section. More importantly, the case of Cand171 (J0917+46) deserves particular attention. 

The WD J0917+46 has $T_{\rm eff}=$ 11288 $\pm$ 72 K and $\log$ g = 5.48 $\pm$ 0.03. The extremely low $\log$ g
is identical to the $\log$ g estimate of the previous lowest mass WD candidate, J2049+00 (Eisenstein et al. 2006). However, Eisenstein et al. find evidence of
multiple minima, and the available photometry for J2049+00 is inconsistent
with this solution (see \S 3). In the case of our lowest gravity object,
J0917+46, both the spectrum and the SED are consistent with the derived
temperature and gravity. Figure 4 presents the $\chi^2$ distribution of our fits to the J0917+46 spectrum, which shows no evidence of any
other minima in the explored parameter space. 

In order to demonstrate that our fit for J0917+46 is reliable, we
present our fits to the flux normalized line profiles in Figure 5. These
fits follow Liebert et al. (2004), Kawka et al. (2005), and Bassa et al. (2006), who
continuum correct their spectra of low mass WDs and fit just the Balmer
lines. We note that the $\chi^2$ (per degree of freedom) is slightly higher in this case (2.08) compared to the $\chi^2$
derived from fitting the entire spectrum (1.45). However, the best fit solution, $T_{\rm eff}=$ 10760 $\pm$ 99 K and $\log$ g =
5.46 $\pm$ 0.04, nicely agrees with our solution presented in Figure 3 and Table 1.
Thus, J0917+46 is the lowest gravity WD currently known.

\section{Understanding The SDSS Low Mass White Dwarf Sample}

The discrepancy between the SDSS photometry and our spectroscopic
analysis initially puzzled us. Even though SDSS photometry of all of our
candidates was consistent with their ELM WD classification, only J0917+46 and J1053+52
turned out to have $\log$ g $<$ 7. What went wrong? We begin by comparing the SEDs of our low
mass candidates with the SDSS broadband photometry.  We then re-analyze
the spectra of other low mass WD candidates previously found in SDSS.

\subsection{Discrepant SDSS Photometry}

We have examined the SEDs of our candidates obtained from the SDSS photometry, and compared them to our
best fit solutions from spectroscopy. We have applied the following AB corrections to the observed SDSS magnitudes: $u-0.04$, $i+0.015$,
$z+0.03$, and no corrections in $g$ and $r$ (see Eisenstein et al. 2006). 
The SDSS database gives the total interstellar absorption and reddening along the line of sight for each star, determined from the dust emission maps of
Schlegel et al. (1998). Since all of our targets are at relatively high Galactic latitudes, the total absorption and reddening are small; the median
total absorption $A_g$ is 0.11, and the median total reddening $E(g - i)$ is 0.05. Therefore, the errors in the reddening corrections
have little effect on our analysis. 

We have calculated the predicted fluxes from our best fit model
atmosphere by integrating over the SDSS filter bandpasses. Without knowing the radii and distances to the WDs, we have used a $\chi^2$
minimization technique to match the predicted fluxes with the observed fluxes. The value of $\chi^2$ is taken as the sum over all
bandpasses of the difference squared between the predicted and observed fluxes, properly weighted by the corresponding observational uncertainties.

Figure 6 shows the SEDs of four of the objects observed at the MMT plus the lowest mass WD candidate
identified by Eisenstein et al. (2006). The de-reddened SDSS photometry and the fluxes
predicted for the parameters derived from our spectroscopic analysis are shown
as error bars and circles, respectively. A comparison of the photometry and our best fit solutions shows that the SDSS $g$-band
photometry is systematically discrepant. For example, our best fit solution for
J0745+18 and J1529+33 (top two panels) reproduce $u,r,i$, and $z$ photometry
reasonably well. However, the SDSS $g$ magnitudes are 20-30\% brighter than expected for these two
objects (as well as many other candidates in the MMT sample). 
In order to check if the discrepancy is caused by our small reddening correction, we have also compared the predicted fluxes to the observed
(not-dereddened) SDSS photometry, and found the same discrepancy in the $g$-band.
Brighter $g$-band magnitudes cause the objects to have more positive $u-g$ colors and
more negative $g-r$ colors that are consistent with low mass WDs. This suggests that SDSS photometry cannot be used by
itself to identify ELM WDs.

We note that the $g$-band magnitude for the previous lowest mass WD
candidate, J2049+00 (Eisenstein et al. 2006), may also be suspect. The bottom
panel of Figure 6 shows that the SED of J2049+00 is discrepant with its
$g$-band magnitude, and thus we revisit its identification as an ELM WD.

\subsection{The SDSS Sample}

We now bring together the low mass WD candidates previously found in SDSS.
Kleinman et al. (2004) identified 2561 WDs, including several WDs with unusually low surface gravities, in the SDSS DR1 spectroscopy data.
Liebert et al. (2004) performed model atmosphere anlysis of two of these unusual WDs and found them to be ELM WDs.
They have also found several other fainter low mass WD candidates with noisy spectra. Eisenstein et al. (2006) 
study WDs across the entire SDSS DR4 and find 13 ELM WD candidates, including the ELM WD candidates identified in
the SDSS DR1.

Eisenstein et al. (2006) use the same model atmospheres as we do, however the $\log$ g sampling of their grid is coarser than ours
(0.50 dex sampling compared to our  0.25 dex sampling). 
The procedure employed by Eisenstein et al. (see Kleinman et al. 2004) makes use
of quadratic interpolation in $\chi^2$ to find the location of the minimum with subpixel resolution, while
our method performs quadratic interpolation in the model fluxes, which is more accurate.
In addition to using the SDSS spectra from 3900 \AA\ to 6800 \AA, Eisenstein et al. also
include SDSS colors in their fits. Using photometry can help constrain the temperature in the spectroscopic analysis of
stars with reliable photometry. However, as we have demonstrated in the previous section, it may not be
ideal to analyze the outliers in color-color diagrams for which the SDSS
photometry may be inaccurate. The lowest mass WD candidate found by Eisenstein et al. (2006) is an
unfortunate example of this.

In order to confirm the ELM WD candidates found by Eisenstein et al. (2006), we use the same procedures for our MMT spectra to fit the SDSS
sample. The SDSS spectra are noisier compared to our MMT data, and the derived parameters are therefore less reliable.
We smooth the spectra to $R=$ 1000 in order to increase the $S/N$ and the models are also degraded to the same
resolution. We fit WD model atmospheres to the
SDSS spectra in the spectral range 3800 \AA\ -- 4500 \AA. The SDSS spectra are usually noisier in the red, and even including H$\beta$
does not change our results significantly. Our temperature estimates differ significantly from the Eisenstein et al. (2006) estimates only for two WDs;
J0822+27 and J1630+42. Using only spectroscopy led to best-fit solutions significantly different from the photometric estimates, and including the $g-r$ colors
in our fits solved the problem for these two stars. The spectroscopic and photometric solutions agreed well for the rest of the SDSS sample.
Figure 7 shows our best fit models (red lines) to the observed SDSS spectra (black lines)
of low mass WD candidates. Table 2 presents $T_{\rm eff}$ and $\log$ g estimates from our study, and those from Liebert et al.
(2004) and Eisenstein et al. (2006). 

Comparisons of temperatures and surface gravities from our analysis, Liebert et al. (2004; squares), and Eisenstein et al. (2006; circles)
are presented in Figure 8. Open symbols represent the objects with low $S/N$ SDSS spectroscopy.
This figure shows that our results for the two ELM WDs analyzed by Liebert et al. (2004) are consistent with their analysis.
Our temperature estimates differ from Eisenstein et al. (2006) results by 141 $\pm$ 1520 K and our surface gravity estimates are different by
0.13 $\pm$ 0.16 dex (excluding the two lowest gravity candidates).
We note that one object, J1053+52, is common to both the MMT and SDSS samples. 
Our fits to the MMT spectrum of this object yield $T_{\rm eff}=$ 15882 $\pm$ 41 K and $\log$ g = 6.40 $\pm$ 0.01, and the
SDSS spectrum yields $T_{\rm eff}=$ 18328 $\pm$ 235 K and $\log$ g = 6.40 $\pm$ 0.05. The nature of this object as an ELM WD is confirmed
by our analysis.

One major difference between this work and Eisenstein et al. is that we find a best fit solution of $\log$ g $\leq$ 5 for their lowest mass
WD candidate, J2049+00. We have repeated our fits for this star by fitting the flux calibrated line profiles
and also fitting the entire SDSS spectrum, and in both cases we find that the best fit solution is exactly our lowest gravity
model ($\log$ g = 5). Eisenstein et al. (2006) find evidence for
multiple minima in their fits for this star, and they conclude that the available photometry
is more consistent with a lower gravity ($\log$ g $<$ 5) solution. Here we demonstrate that the SDSS spectrum of J2049+00 is also
consistent with a lower gravity ($\log$ g $<$ 5) solution, and therefore it is not a WD. We performed the Clewley et al. (2002) line shape test for the
classification of halo A-type stars, but the results are ambiguous. The H$\gamma$ profile is marginally consistent with the star being a
blue horizontal branch (BHB) star, but the H$\delta$ profile is inconsistent with a BHB star.
Based on H line indices we determine the spectral type of J2049+00 to be A2 with an uncertainty of $\pm$ 1.2 sub-types.
If J2049+00 is an A star with $M_{\rm V}\sim$ 1, then it is located at $\sim$ 36 kpc.
The lack of a significant proper motion in the USNO-B catalog is consistent with it being a distant A star.

Our fits also indicate that J0849+04 (with $S/N<$ 10 spectroscopy), Eisenstein et al.'s second lowest mass WD candidate with $T_{\rm eff}=$ 9962 K and $\log$ g = 5.93
$\pm$ 0.15, is better fit with a $T_{\rm eff}=$ 12000 $\pm$ 208 K and $\log$ g = 7.34 $\pm$ 0.07 model atmosphere. Our fits to the Balmer
lines (continuum corrected spectrum) also result in a best fit solution of $T_{\rm eff}=$ 12397 $\pm$ 91 K and $\log$ g = 7.21 $\pm$ 0.05.
Both of these fits suggest that J0849+04 is a low mass WD, but not an ELM WD.

The rest of the objects in the Eisenstein et al. low mass WD sample seem to have surface gravities in the range $\log$ g = 6 -- 7.3,
and hence our analysis supports their classification as low mass WDs. However, the SDSS spectroscopy for the majority of them is very noisy,
especially in the blue, and a better understanding of these objects must await higher $S/N$ spectroscopy.

\section{Discussion}

Our spectroscopic analysis of the low mass WD candidates in the SDSS shows that 
J0917+46 has $T_{\rm eff}=$ 11,288 $\pm$ 72 K and $\log$ g = 5.48 $\pm$ 0.03, and
it is the lowest surface gravity WD currently known. We find one other ELM WD in our MMT sample, J1053+52.
The remaining 40 candidates observed at the MMT are normal DA WDs with apparently inaccurate SDSS $g$ magnitudes.
We also confirm the low surface gravities of the ELM WD candidates identified by Eisenstein et al. (2006), with the exception
of their lowest mass WD candidate.

Figure 9 shows the effective temperatures and surface gravities for the MMT (filled circles) and SDSS (triangles) samples plus the previously
identified ELM WDs in the literature (open circles). Solid lines show the constant mass tracks for low mass WDs\footnote{The
difference between the low mass WD models below and above 0.18$M_\sun$ is caused by element diffusion and thermonuclear flashes.
WDs with $M > 0.18 M_\sun$ suffer from several thermonuclear
flashes which consume most of the H-rich envelope.
Element diffusion induces additional flashes which ultimately leads to thinner H envelopes.
As a result, when the final cooling branch is reached, these WDs will be characterized by a thin hydrogen envelope.
The models with $M < 0.18 M_\sun$ do not experience thermonuclear flashes even if diffusion is considered. As a result,
they have massive hydrogen envelopes and therefore their radius will be considerably larger than the
more massive counterparts (L. Althaus 2006, private communication).}
from Althaus et al. (2001;
labeled in $M_\sun$ on the right side of the figure),
and the tracks for zero age main sequence and horizontal branch stars. The zero-age BHB marks the location of the star at the start
of the core-helium burning phase. Stars evolve upward to lower $\log$ g values as the He burning proceeds.

Figure 9 shows that the majority of the WDs found by Brown et al. (2006) have $M\sim0.5M_\sun$. Only two objects, J0917+46 and J1053+52
have $M<0.2M_\sun$. All of the low mass WD candidates identified by Eisenstein et al. (2006), except J2049+00, have $M
\leq0.36M_\sun$, therefore they should have formed in binary systems. None of the stars in the MMT sample or the SDSS sample show
excesses in their SDSS photometry, therefore the companions are likely to be compact objects. 
We have two separate spectra for J1053+52. We have measured a radial velocity of $-53~ \pm$ 18 km s$^{-1}$ from
the MMT spectrum obtained on 24 Feb 2006, and a velocity of 11 $\pm$ 38 km s$^{-1}$ from the SDSS observations obtained on 10 Jan 2003.
The observed radial velocity change may be real, however more accurate and time-resolved spectroscopic observations of this object,
as well as all the other ELM WDs in our study, are required to constrain the mass of the unseen companion.

Figure 9 demonstrates that J0917+46 is an ELM WD and not a BHB or
a main-sequence star. Its temperature and surface gravity imply an
absolute magnitude of $M_{\rm V}\sim6.9$ (Althaus et al. 2001) and an age of 470 Myr. This
luminosity places it at a distance of 2.6 kpc. At a Galactic latitude of $+44^{\circ}$, it is located at 1.8 kpc above the plane.
J0917+46 has a proper motion measurement of 0 mas/yr in the USNO-B catalog, which is consistent with a distant object.
Our radial velocity measurement of 165 $\pm$ 12 km s$^{-1}$ from the MMT spectroscopy corresponds to U = $-$108 km s$^{-1}$, V = 25 km s$^{-1}$, and
W = 122 km s$^{-1}$ (heliocentric), consistent with a thick disk or halo WD (Chiba \& Beers 2000).
If the observed radial velocity is due to the orbital motion of the star, then it would be more consistent with a thick disk WD.

Low mass WDs can be produced in Low Mass X-ray Binaries (LMXBs; WD - neutron star binaries) or in
a common-envelope phase with other companion stars.
About 90\% of the known LMXBs are located in the Galactic plane ($b\leq 20^\circ$; Ritter \& Kolb 2003). If J0917+46 had a neutron star companion,
we would need an explanation for its peculiar location above the Galactic plane.
The unusually high velocity of another extremely low mass WD, LP400-22 (Kawka et al. 2006), indicates
that LP400-22 may have been released from its close binary. A similar scenario may explain the peculiar location of
J0917+46.

Althaus et al. (2001) models predict J0917+46 to have $M\sim0.17M_\sun$, $L\sim0.2L_\sun$, and $R\sim0.12R_\sun$, about 9 times bigger than a typical
0.6$M_\sun$ WD. Other calculations and mass radius relations should give similar results, within a few 0.01 $M_\sun$ (see Bassa et al. 2006).
A close examination of the spectrum of this object reveals a Ca K line (Cand171 in Figure 3). The Ca K line has an equivalent
width of 0.35 \AA\ and a radial velocity measurement of 164 km s$^{-1}$; it is photospheric. The equivalent width of the Ca line is similar to the
DAZs observed by Zuckerman et al. (2003) and Koester et al. (2005), and corresponds to an abundance of $\log$ (Ca/H) = $-$5.89 (nearly solar).
J0917+46 has more Ca than many of the DAZs with circumstellar debris disks, and requires an external source for the observed metals (Kilic et al. 2006;
Kilic \& Redfield 2007). The star is located far above the Galactic plane where accretion from the ISM is unlikely (Dupuis et al. 1993). 
The nature of the possible companion star needs to be determined before the Ca abundance can be explained.

Extremely low mass WDs seem to be rare. Eisenstein et al. (2006) found only 13 candidates in the SDSS DR4 area (4783 deg$^2$). We
have identified 2 ELM WDs (one in common with their analysis) in 3000 deg$^2$. However, our search was limited to $-0.39 <g-r< -0.27$
(and therefore to $T_{\rm eff}>$ 11,000 K for $\log$ g $<$ 6 WDs). Eisenstein et al. (2006) made an initial color cut to select their WD
candidates for spectroscopic analysis. Their initial color cut included the $u-g$ vs. $g-r$ region where we expect to find $\log$ g $\geq$
5 WDs, however they did not find any WD candidates with $M<$ 0.17 $M_\sun$. 

Detailed calculations of the evolution of 1-3.5 $M_\sun$ stars
in close binary systems with neutron stars show that the final mass of the He-core WD produced in the process can be as low as 0.02 $M_\sun$
(Benvenuto \& De Vito 2005). Ergma et al. (1998) argued that if the initial orbital period of a 1$M_\sun$ population
II object in a binary system with a neutron star is less than 0.95 days, the orbital evolution of the system would proceed towards
very short orbital periods, and these systems would end their evolution as ultra-compact LMXBs with a WD secondary mass of 
$\leq 0.1 M_\sun$ (e.g. 4U 1820-30 is an LMXB with an orbital period of 11 minutes and an estimated secondary mass of $M\sim 0.06-0.08M_\sun$; 
Stella et al. 1987; Rappaport et al. 1987). They also argued that if the initial orbital period of the system is between 0.95 and $\sim$1.50 days
(the upper limit is the bifurcation orbital period), the systems would evolve towards shorter orbital periods and LMXBs with orbital periods
of $\geq$ 10 hours will be formed. Ergma et al. (1998) found that the lowest mass WDs in such systems would be 0.15-0.16 $M_\sun$ and 0.17$M_\sun$ for
population I and II stars, respectively. These low mass limits also seem consistent with the analysis of Benvenuto \& De Vito (2005). 

Binary millisecond pulsar systems with low mass WD secondaries are thought to be the descendants of the LMXBs.
All of the currently known WDs in the field and around the millisecond pulsars have masses larger than 0.17 $M_\sun$.
The masses of some of these WDs have been determined from the Shapiro (1964) delay of the pulsar signal in the gravitational field
of the companion WD (L\"ohmer et al. 2005). This delay is linearly proportional to the mass of the companion, and therefore the non-detection of
$M<$ 0.17 $M_\sun$ WDs around pulsars may be an observational bias. A more targeted search for these objects may find WDs with
smaller masses, however identifying these low gravity objects as low mass WDs may be challenging.
On the other hand, if the binary formation models of low mass He-core WDs around neutron stars are correct, we would expect to find lower mass WDs
only in ultra-compact LMXBs and their descendants. 

\section{Conclusions}

We have performed spectroscopic analysis of the low mass WD candidates found in the SDSS. We have shown that 
SDSS J0917+46 is the lowest mass WD currently known with $M \sim 0.17 M_\sun$. 
In addition, there are three more WDs in the SDSS with masses smaller than 0.2 $M_\sun$. These WDs presumably have unseen companions.
A radial velocity search for these unseen companions will be necessary to constrain their masses and the evolutionary
scenarios for the formation of extremely low mass WDs. So far both of the spectroscopically
confirmed $\sim$0.2 $M_\sun$ WDs with known companions (PSR J1012+5307 and J1911-5958A) are in 14.5 - 20 hour orbits around neutron star companions
(van Kerkwijk et al. 1996; Bassa et al. 2006). If J0917+46 is in a binary system with a neutron star, we expect a similar orbital period.

The search for radio pulsars around 8 low mass WDs by van Leeuwen et al. (2006) did not find any pulsar companions, and they concluded
that the fraction of low mass He-core WDs with neutron star companions is less than 18\% $\pm$ 5\%. However, their sample included only
one WD with $M\sim 0.2 M_\sun$ (SDSS J1234-02).
Liebert et al. (2004) argued that if the binary separation is appropriate, it is possible to create an extremely low mass WD
in a WD - WD binary as well. Time-resolved radial velocity measurements will be necessary to differentiate between these two formation scenarios.
Excluding the ultra-short period LMXBs, the models for the formation of low mass helium WDs in millisecond pulsar binary systems predict the lowest
mass WDs to have $M \sim 0.17M_\sun$. J0917+46 may be an example of the lowest mass WDs produced in such systems.

\acknowledgements
We would like to thank Marc Pinsonneault, Gregory Sivakoff, and Andrew Gould for helpful discussions.
We also thank our anonymous referee for useful suggestions.

\clearpage
\begin{deluxetable}{llccrrrlcc}
\tabletypesize{\scriptsize}
\tablecolumns{10}
\tablewidth{0pt}
\tablecaption{Low Mass White Dwarf Candidates Observed at the MMT}
\tablehead{
\colhead{Name}&
\colhead{SDSS J}&
\colhead{$g$}&
\colhead{$u-g$}&
\colhead{$g-r$}&
\colhead{$r-i$}&
\colhead{$i-z$}&
\colhead{$T_{\rm eff}$ (K)}&
\colhead{$\log$ g}&
\colhead{$\chi^2$}
}
\startdata
Cand017 & 145859.60+431113.3 & 19.201 & 0.539 & --0.324 & --0.228 & --0.320 & 13000 $\pm$ 319 & 7.72 $\pm$ 0.04 & 1.95\\
Cand020 & 151027.67+412520.1 & 19.008 & 0.484 & --0.351 & --0.262 & --0.478 & 12420 $\pm$ 128 & 7.58 $\pm$ 0.02 & 2.53\\
Cand025 & 152905.82+330520.4 & 18.288 & 0.575 & --0.331 & --0.185 & --0.142 & 11519 $\pm$ 66 & 8.11 $\pm$ 0.02 & 1.18\\
Cand036 & 155708.48+282336.0 & 17.496 & 0.532 & --0.366 & --0.223 & --0.211 & 12044 $\pm$ 65 & 7.74 $\pm$ 0.02 & 1.13\\
Cand102 & 002803.34$-$001213.4 & 18.346 & 0.516 & --0.417 & --0.263 & --0.237 & 14788 $\pm$ 61 & 7.86 $\pm$ 0.02 & 1.36\\
Cand106 & 010044.69$-$005034.1 & 19.981 & 0.578 & --0.301 & --0.113 & --0.032 & 16248 $\pm$ 139 & 7.39 $\pm$ 0.05 & 1.35\\
Cand107 & 010657.83$-$100839.3 & 19.287 & 0.526 & --0.367 & --0.168 & --0.089 & 15263 $\pm$ 68 & 7.71 $\pm$ 0.02 & 1.36\\
Cand108 & 011130.67+141049.8 & 19.818 & 0.380 & --0.358 & --0.236 & --0.309 & 17426 $\pm$ 84 & 7.88 $\pm$ 0.02 & 1.34\\
Cand116 & 020232.30$-$084918.3 & 19.117 & 0.482 & --0.347 & --0.125 & --0.151 & 10678 $\pm$ 47 & 8.56 $\pm$ 0.02 & 1.20\\
Cand124 & 031240.49$-$005941.1 & 19.367 & 0.283 & --0.411 & --0.417 & --0.192 & 17251 $\pm$ 50 & 7.70 $\pm$ 0.01 & 1.63\\
Cand128 & 074508.15+182630.0 & 19.295 & 0.750 & --0.403 & --0.195 & --0.140 & 11000 $\pm$ 43 & 8.23 $\pm$ 0.01 & 1.27\\
Cand133 & 075637.74+203730.7 & 18.578 & 0.229 & --0.415 & --0.297 & --0.497 & 17177 $\pm$ 61 & 7.87 $\pm$ 0.02 & 1.44\\
Cand137 & 080234.62+432301.1 & 18.356 & 0.432 & --0.347 & --0.264 & --0.324 & 13458 $\pm$ 111 & 7.76 $\pm$ 0.01 & 1.40\\
Cand144 & 082003.03+250012.1 & 19.454 & 0.449 & --0.381 & --0.093 & --0.333 & 12844 $\pm$ 84 & 7.94 $\pm$ 0.01 & 1.28\\
Cand150 & 083303.03+365906.3 & 17.930 & 0.947 & --0.283 & --0.283 & --0.229 & 13399 $\pm$ 120 & 7.76 $\pm$ 0.02 & 1.31\\
Cand164 & 085652.65+233341.8 & 18.389 & 0.311 & --0.401 & --0.163 & --0.331 & 14255 $\pm$ 55 & 7.80 $\pm$ 0.01 & 1.22\\
Cand171 & 091709.55+463821.8 & 18.696 & 0.738 & --0.328 & --0.237 & --0.219 & 11288 $\pm$ 72 & 5.48 $\pm$ 0.03 & 1.45\\
Cand176 & 092918.13+374529.9 & 19.325 & 0.470 & --0.377 & --0.287 & --0.160 & 16480 $\pm$ 55 & 7.51 $\pm$ 0.02 & 1.07\\
Cand186 & 095030.48+385713.2 & 18.347 & 0.322 & --0.401 & --0.314 & --0.179 & 17788 $\pm$ 45 & 7.79 $\pm$ 0.01 & 1.02\\
Cand192 & 095641.27+685727.5 & 18.056 & 0.550 & --0.372 & --0.364 & --0.346 & 13000 $\pm$ 127 & 7.75 $\pm$ 0.02 & 5.18\\
Cand193 & 095717.31+552839.2 & 18.678 & 0.352 & --0.388 & --0.337 & --0.334 & 16381 $\pm$ 57 & 7.76 $\pm$ 0.02 & 1.28\\
Cand203 & 100929.84+411205.2 & 19.424 & 0.404 & --0.348 & --0.233 & --0.186 & 14621 $\pm$ 90 & 7.93 $\pm$ 0.02 & 1.22\\
Cand205 & 101519.62+595430.5 & 17.920 & 0.544 & --0.309 & --0.181 & --0.159 & 14965 $\pm$ 71 & 7.82 $\pm$ 0.03 & 1.05\\
Cand207 & 101557.01+510428.4 & 18.707 & 0.575 & --0.348 & --0.191 & --0.102 & 16316 $\pm$ 59 & 7.84 $\pm$ 0.02 & 1.36\\
Cand208 & 101647.50+375835.2 & 17.279 & 0.375 & --0.402 & --0.303 & --0.299 & 16908 $\pm$ 37 & 7.81 $\pm$ 0.01 & 1.17\\
Cand225 & 104650.53+510028.6 & 17.867 & 0.418 & --0.400 & --0.269 & --0.273 & 15866 $\pm$ 61 & 7.88 $\pm$ 0.02 & 1.29\\
Cand352 & 073734.88+215017.3 & 19.208 & 0.459 & --0.320 & --0.088 & --0.432 & 12138 $\pm$ 56 & 7.92 $\pm$ 0.01 & 1.10\\
Cand362 & 082542.49+080100.5 & 19.342 & 0.451 & --0.366 & --0.313 & --0.347 & 14087 $\pm$ 95 & 7.83 $\pm$ 0.02 & 1.13\\
Cand363 & 082823.55+470001.4 & 17.634 & 0.423 & --0.321 & --0.188 & --0.446 & 14526 $\pm$ 58 & 7.67 $\pm$ 0.02 & 1.40\\
Cand369 & 084823.75+050845.2 & 17.027 & 0.271 & --0.376 & --0.256 & --0.287 & 15332 $\pm$ 38 & 7.18 $\pm$ 0.01 & 1.21\\
Cand371 & 085141.58+425117.7 & 18.267 & 0.493 & --0.304 & --0.248 & --0.147 & 12337 $\pm$ 57 & 7.92 $\pm$ 0.01 & 1.15\\
Cand375 & 085859.83+050757.2 & 19.018 & 0.396 & --0.357 & --0.244 & --0.142 & 13464 $\pm$ 97 & 7.48 $\pm$ 0.02 & 0.98\\
Cand386 & 092001.94+005201.2 & 19.288 & 0.441 & --0.373 & --0.187 & --0.141 & 13170 $\pm$ 145 & 7.78 $\pm$ 0.02 & 1.43\\
Cand389 & 093955.29+054934.6 & 18.178 & 0.539 & --0.344 & --0.214 & --0.278 & 12877 $\pm$ 90 & 7.68 $\pm$ 0.01 & 1.58\\
Cand399 & 100354.41+063850.6 & 19.115 & 0.416 & --0.336 & --0.102 & --0.239 & 14351 $\pm$ 62 & 7.81 $\pm$ 0.01 & 1.14\\
Cand400 & 100618.54+605500.7 & 18.155 & 0.551 & --0.283 & --0.204 & --0.105 & 14942 $\pm$ 55 & 7.70 $\pm$ 0.01 & 1.15\\
Cand426 & 105353.89+520031.0 & 18.873 & 0.396 & --0.358 & --0.235 & --0.429 & 15882 $\pm$ 41 & 6.40 $\pm$ 0.01 & 1.32\\
Cand430 & 105850.47+432710.9 & 18.936 & 0.450 & --0.372 & --0.252 & --0.097 & 13600 $\pm$ 43 & 7.35 $\pm$ 0.01 & 1.34\\
Cand433 & 110200.07$-$004055.9 & 19.103 & 0.332 & --0.364 & --0.243 & --0.149 & 14102 $\pm$ 127 & 7.94 $\pm$ 0.02 & 1.18\\
Cand455 & 111908.64+033818.6 & 18.857 & 0.403 & --0.338 & --0.210 & --0.212 & 13837 $\pm$ 113 & 7.81 $\pm$ 0.02 & 1.23\\
Cand477 & 113807.13+143257.4 & 17.619 & 0.596 & --0.283 & --0.265 & --0.220 & 12514 $\pm$ 76 & 7.24 $\pm$ 0.01 & 1.42\\
Cand487 & 114344.96+052214.8 & 17.364 & 0.441 & --0.358 & --0.354 & --0.178 & 14243 $\pm$ 55 & 7.73 $\pm$ 0.01 & 1.25\\
\enddata
\end{deluxetable}

\clearpage
\begin{deluxetable}{llclcr}
\tabletypesize{\scriptsize}
\tablecolumns{10}
\tablewidth{0pt}
\tablecaption{Low Mass White Dwarf Candidates Identified by Eisenstein et al. (2006)}
\tablehead{
\colhead{Object}&
\colhead{$T_{\rm eff}$ (K; this work)}&
\colhead{$\log$ g (this work)}&
\colhead{$T_{\rm eff}$ (K; SDSS)}&
\colhead{$\log$ g (SDSS)}&
\colhead{$S/N^a$}
}
\startdata
J0022+00 & 19562 $\pm$ 153 & 7.12 $\pm$ 0.05 & 17355 $\pm$ 394 & 6.95 $\pm$ 0.11 & 9.6\\
J0022$-$10 & 18397 $\pm$ 299 & 6.82 $\pm$ 0.07 & 19444 $\pm$ 758 & 6.76 $\pm$ 0.16 & 7.0\\
J0822+27 & 8937 $\pm$ 48 & 7.21 $\pm$ 0.04 & 8777 $\pm$ 40 & 6.78 $\pm$ 0.11 & 20.3\\
J0849+04 & 12000 $\pm$ 208 & 7.34 $\pm$ 0.07 & 9962 $\pm$ 165 & 5.93 $\pm$ 0.15 & 8.5\\
J1053+52 & 18328 $\pm$ 235 & 6.40 $\pm$ 0.05 & 15399 $\pm$ 400 & 6.28 $\pm$ 0.11 & 10.7\\
J1056+65 & 19000 $\pm$ 151 & 7.03 $\pm$ 0.05 & 20112 $\pm$ 634 & 6.94 $\pm$ 0.12 & 9.4\\
J1056+65$^b$ &  &  & 21910 $\pm$ 1900 & 7.07 $\pm$ 0.10 & \\
J1234$-$02 & 16897 $\pm$ 78 & 6.49 $\pm$ 0.02 & 17114 $\pm$ 227 & 6.30 $\pm$ 0.05 & 21.2\\
J1234$-$02$^b$ &  &  & 17470 $\pm$ 750 & 6.38 $\pm$ 0.05 & \\
J1426+01 & 15869 $\pm$ 112 & 7.02 $\pm$ 0.03 & 16311 $\pm$ 359 & 6.92 $\pm$ 0.09 & 10.7\\
J1436+50 & 18339 $\pm$ 110 & 6.59 $\pm$ 0.02 & 16993 $\pm$ 229 & 6.58 $\pm$ 0.06 & 17.6\\
J1625+36 & 23000 $\pm$ 435 & 6.01 $\pm$ 0.07 & 24913 $\pm$ 936 & 6.20 $\pm$ 0.15 & 8.7\\
J1630+42 & 14183 $\pm$ 942 & 7.08 $\pm$ 0.07 & 14854 $\pm$ 359 & 6.89 $\pm$ 0.13 & 7.0\\
J2049+00 & 8581 $\pm$ 31 & 5.00 $\pm$ 0.12 & 8660 $\pm$ 144 & 5.48 $\pm$ 0.10 & 12.9\\
J2252$-$00 & 19114 $\pm$ 99 & 7.12 $\pm$ 0.03 & 20479 $\pm$ 433 & 6.85 $\pm$ 0.08 & 13.5\\
\enddata
\tablecomments{$(a)$ $S/N$ of the SDSS spectra in the $g$-band. $(b)$ $T_{\rm eff}$ and $\log$ g fits by Liebert et al. (2004).}
\end{deluxetable}

\clearpage

\begin{figure}
\plotone{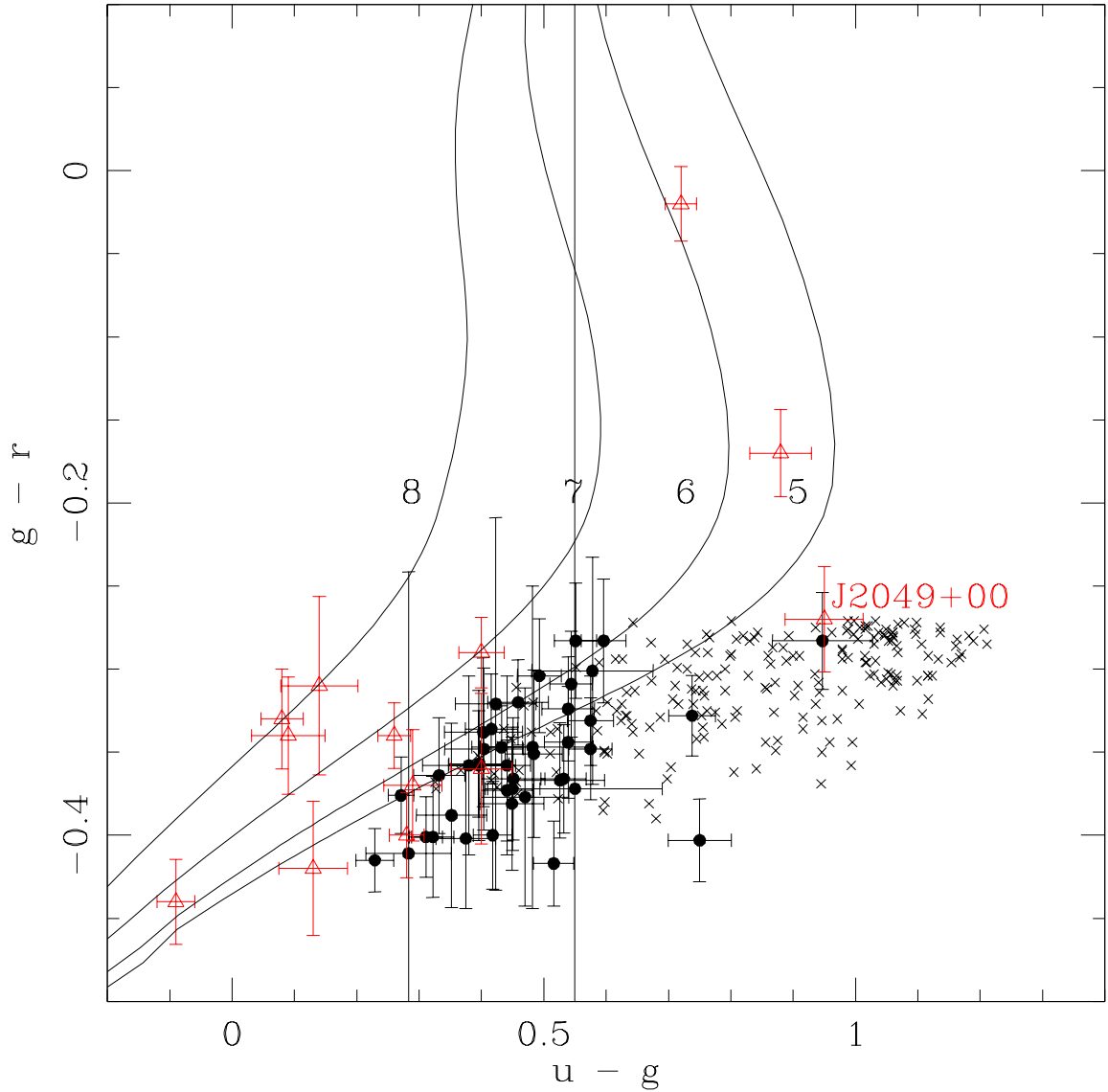}                
\caption{Color-color diagram for the low mass WD candidates from Brown et al. (2006; circles) and Eisenstein et al. (2006;
triangles). The lowest mass WD candidate found by Eisenstein et al. (2006), SDSS J2049+00, is labeled.
Our synthetic photometry of WD model atmospheres with $\log$ g = 8, 7, 6, and 5 (Finley et al. 1997) are shown as solid
lines. Late B stars found in Brown et al.'s (2006) survey are shown as crosses.}
\end{figure}

\begin{figure}
\plotone{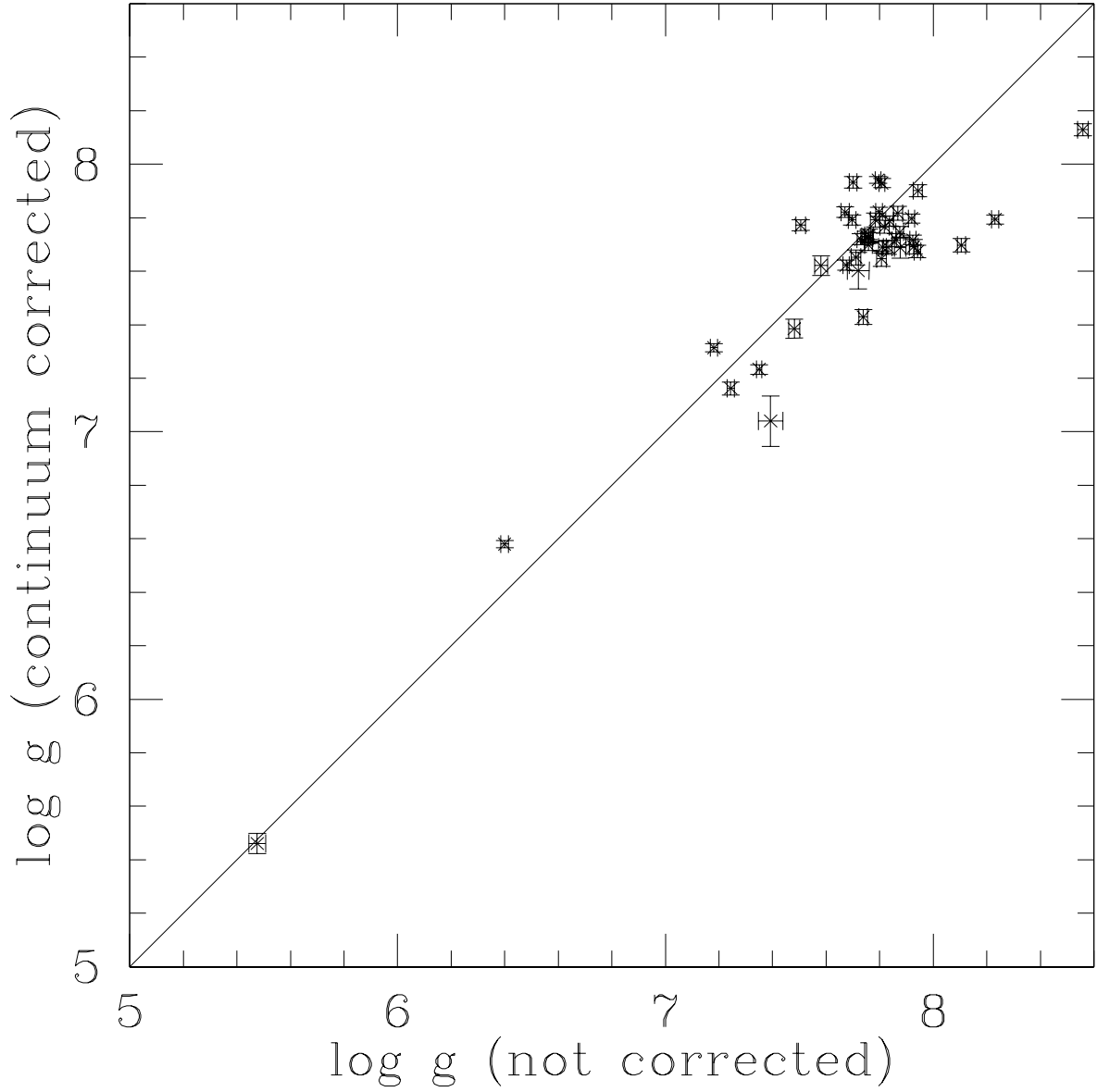}              
\caption{$\log$ g determinations using the line profiles from the continuum corrected spectra versus using the observed spectra without any continuum correction.}
\end{figure}

\begin{figure}
\plotone{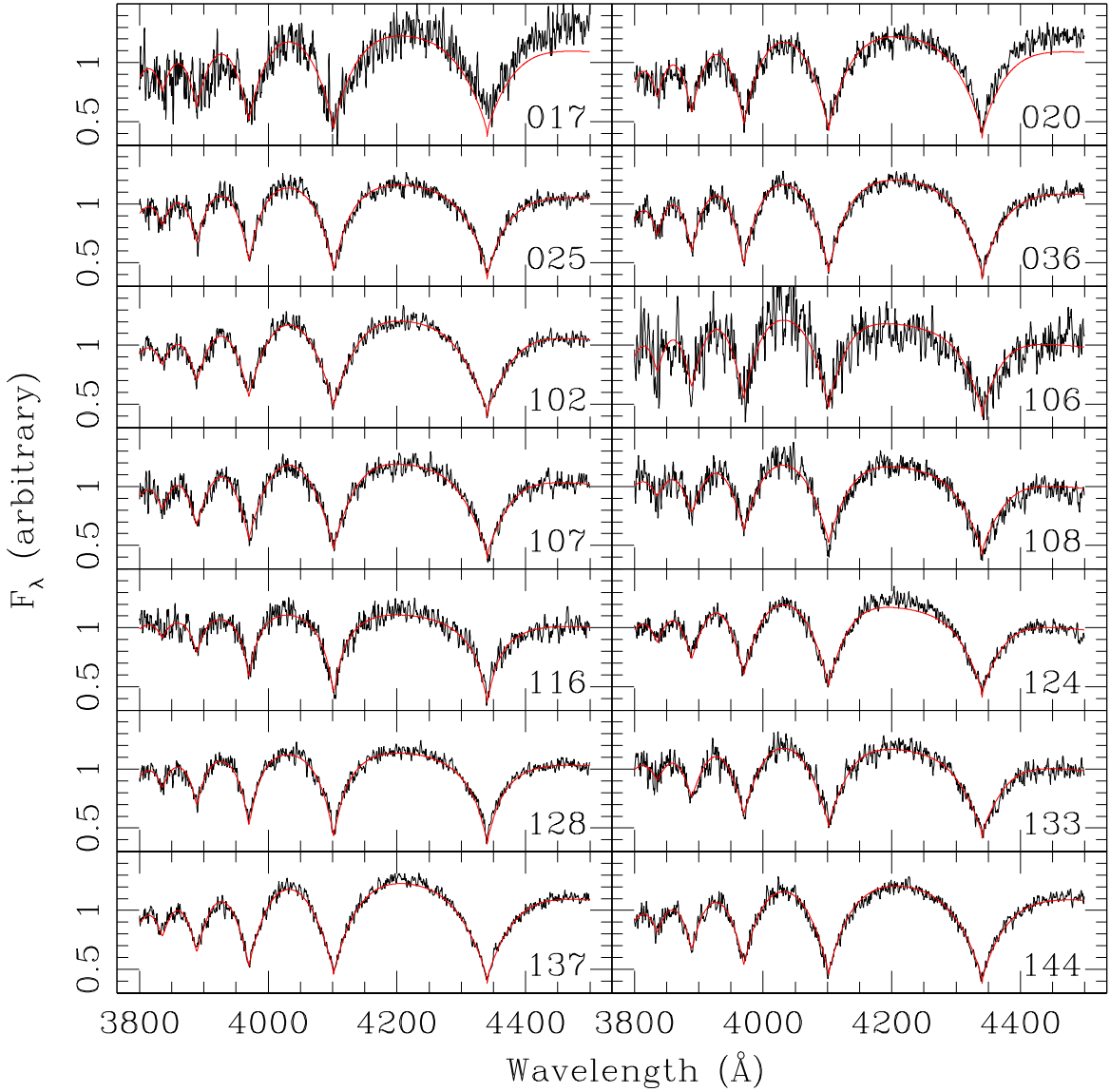}                
\caption{Spectral fits (red lines) to the MMT spectra (black lines) of low mass WD candidates. Candidate names are given in the lower right corner of each panel.}
\end{figure}
\clearpage
{\plotone{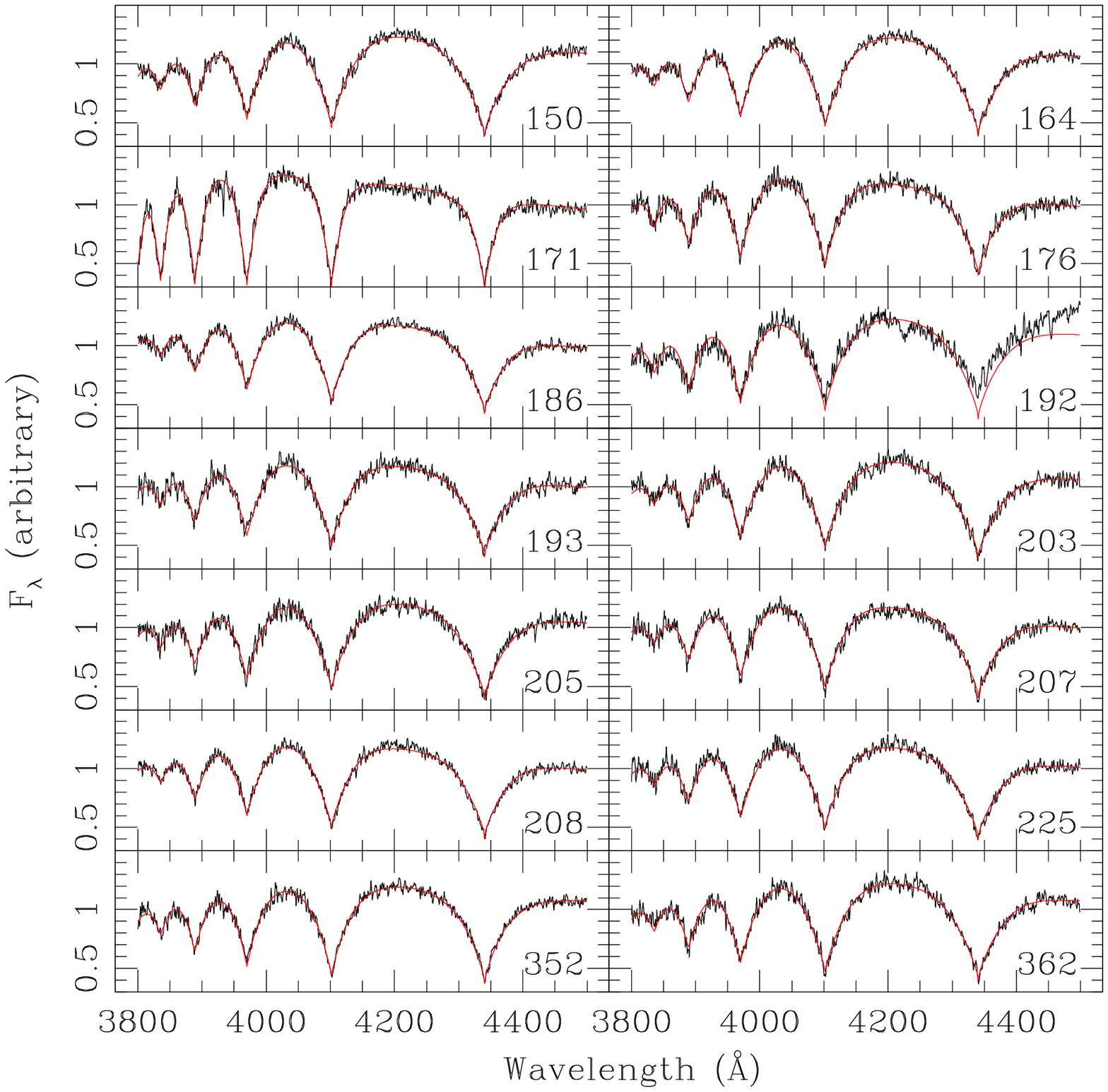}}                
\centerline{Fig. 3. --- continued.}
\clearpage
{\plotone{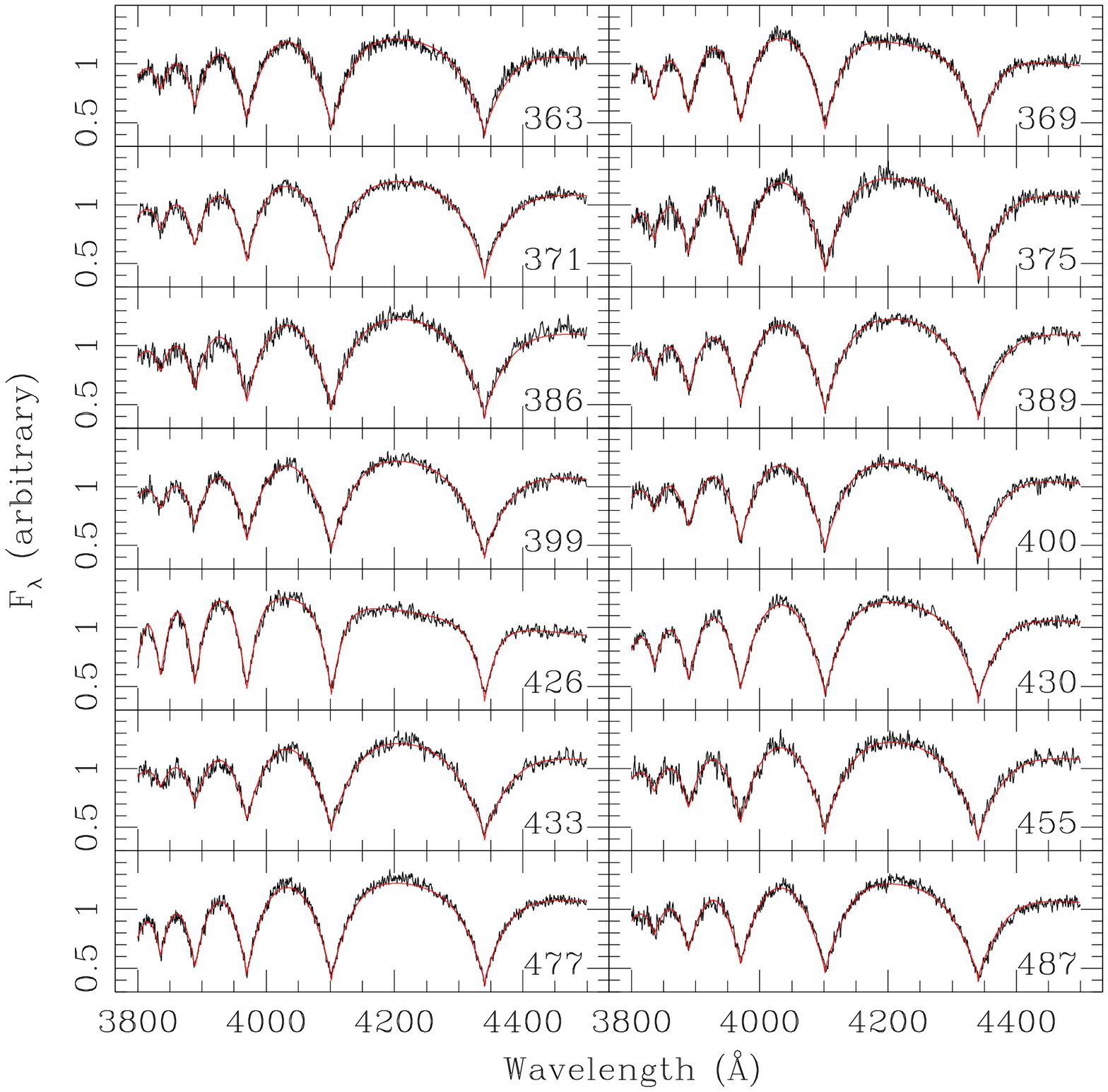}}                
\centerline{Fig. 3. --- continued.}

\begin{figure}
\includegraphics[angle=90,scale=.7]{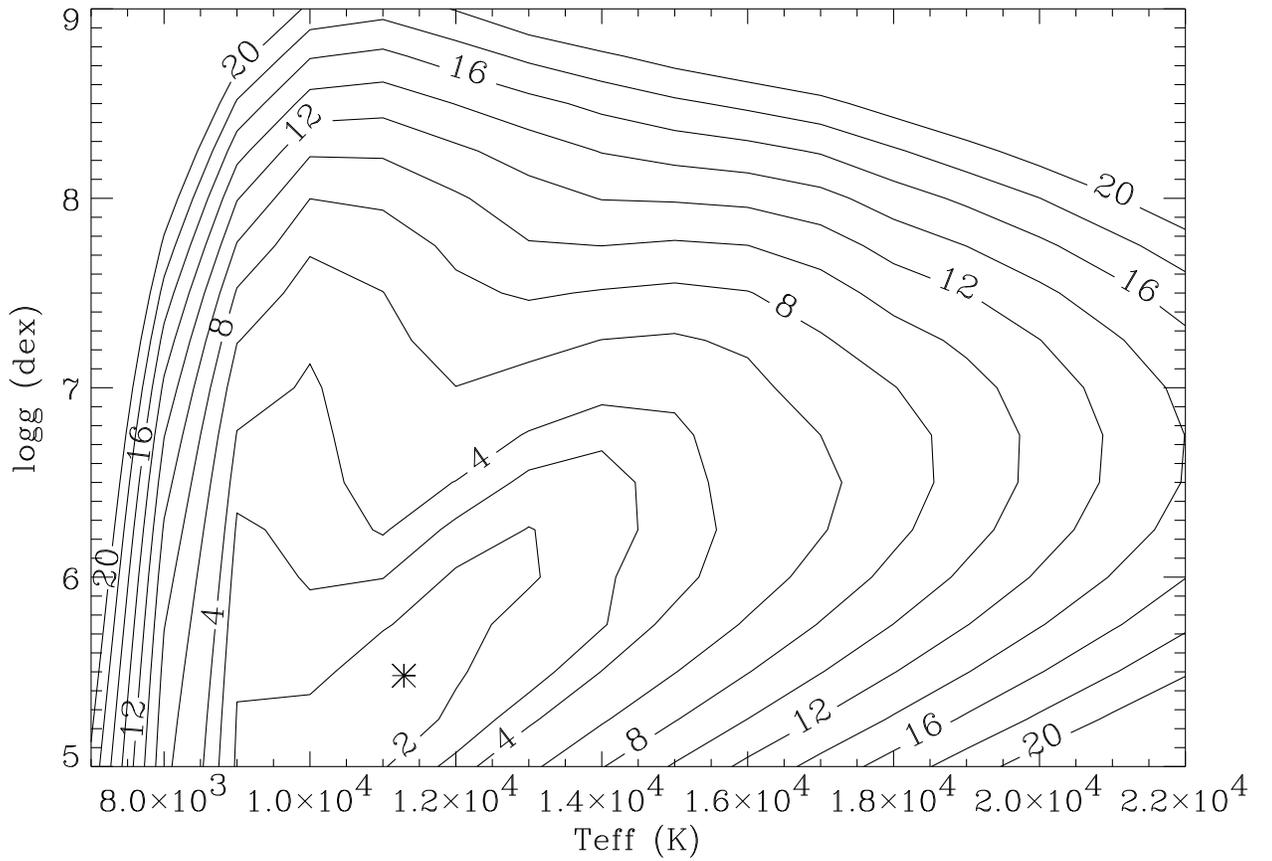}
\caption{Contours of the $\chi^2$ (per degree of freedom) distribution of our fits to the spectrum of the ELM WD J0917+46. An asterisk marks the location of the best fit solution with a minimum $\chi^2=$ 1.45.}
\end{figure}

\begin{figure}
\plotone{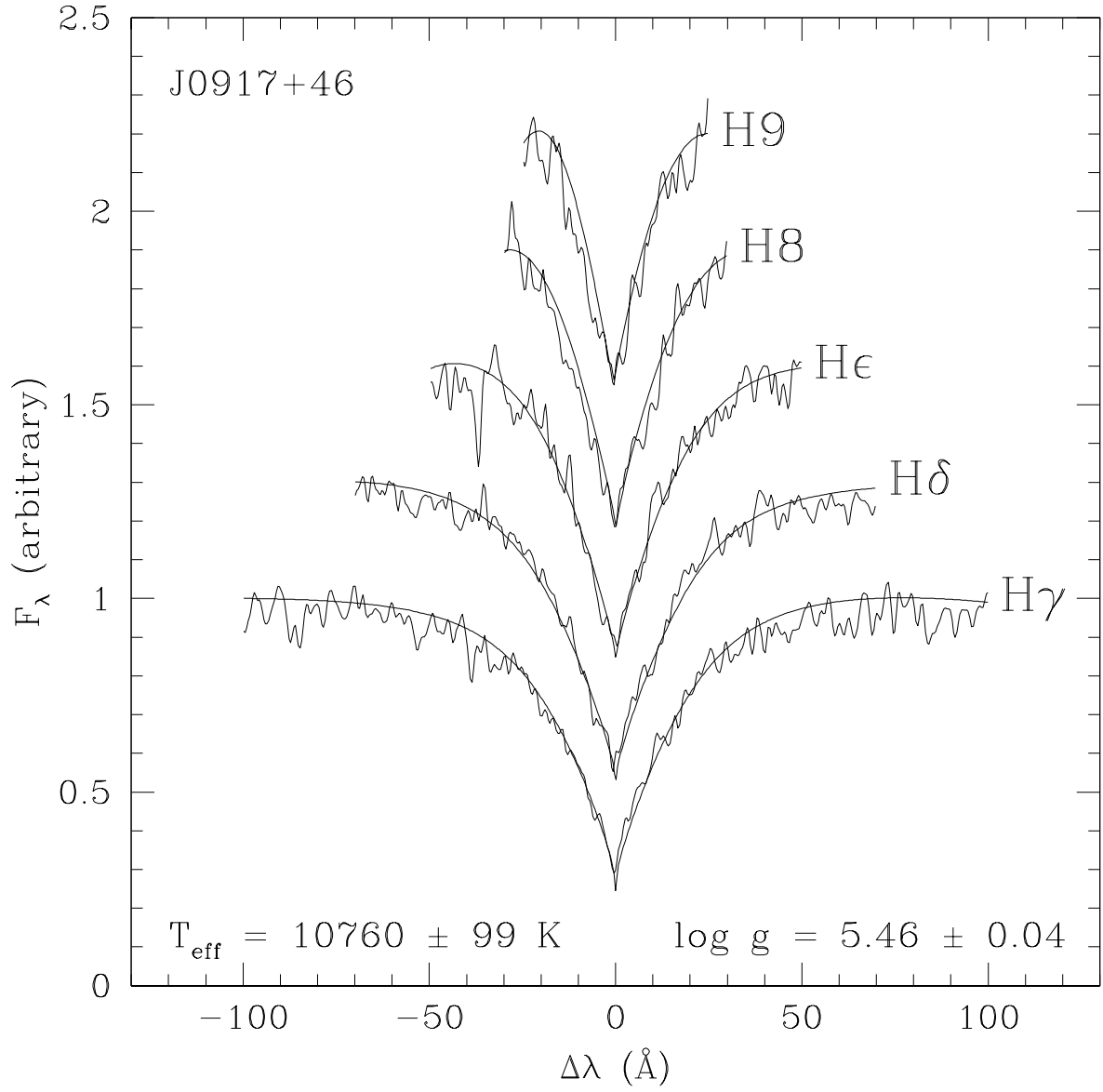}                
\caption{Spectral fits to the flux-normalized line profiles of J0917+46.}
\end{figure}

\begin{figure}
\plotone{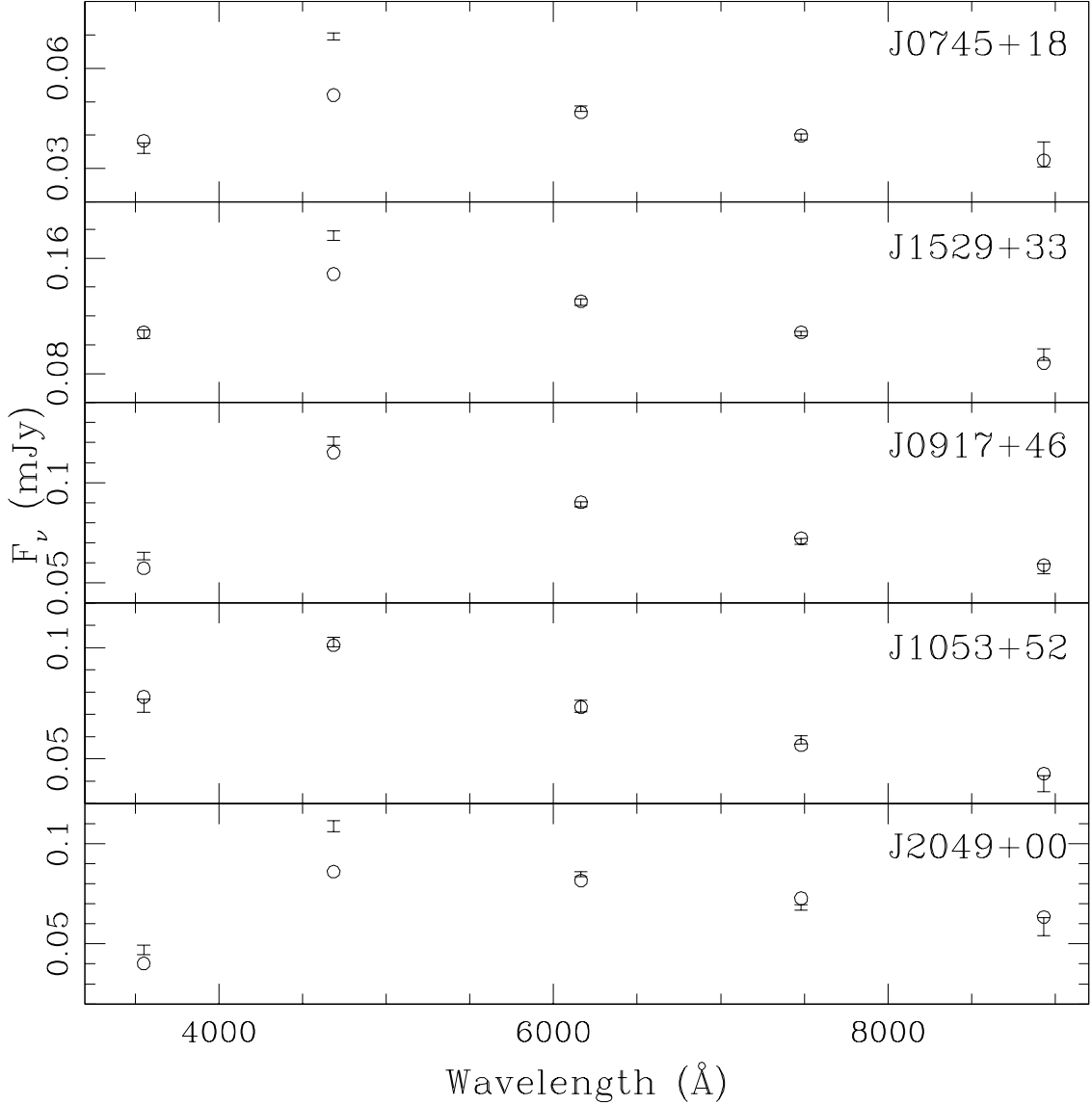}                
\caption{Spectral energy distributions of four objects from our study and one object from Eisenstein et al. (2006). The SDSS photometry
and the predicted fluxes from our best fit solution to the spectra are shown as error bars and circles, respectively. The SDSS photometry and the
best fit solution for J2049+00 is taken from Eisenstein et al. (2006). J0745+18 and J1529+33 have $\log$ g $>$ 8 and they are good examples of
objects with discrepant $g$-band photometry, which causes the colors for these objects to appear similar to low mass WDs. J0917+46
and J1053+52 are true low mass WDs with reliable SDSS photometry. The $g$-band photometry for J2049+00, the lowest mass WD
candidate identified by Eisenstein et al. (2006), is also suspect, which suggests that it is not located in the $u-g$ vs.
$g-r$ color-color region where we expect to find low mass WDs.}
\end{figure}

\begin{figure}
\plotone{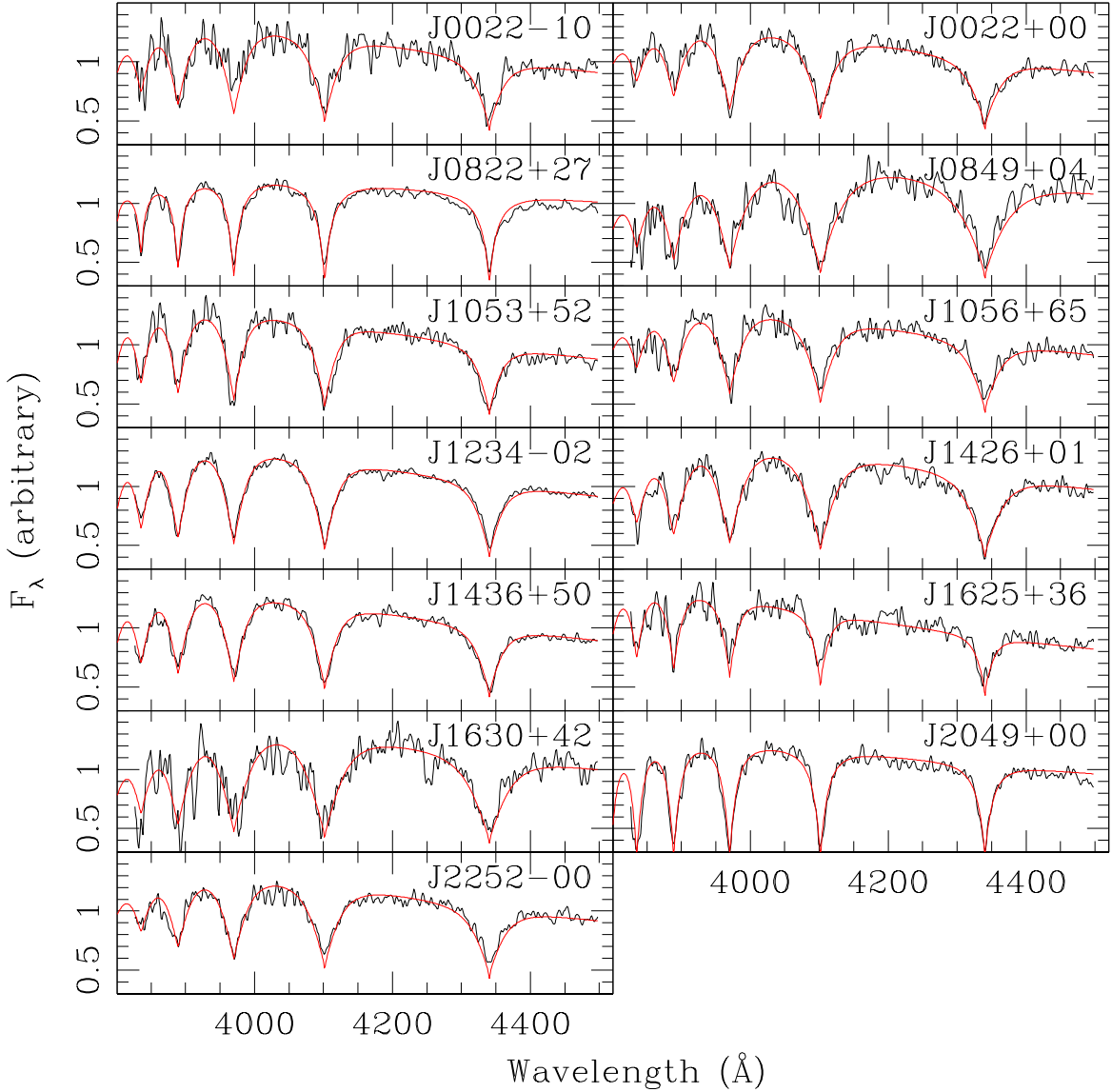}                
\caption{Spectral fits (red lines) to the SDSS spectra (black lines) of low mass WD candidates identified by Eisenstein et al. (2006).}
\end{figure}

\begin{figure}
\plotone{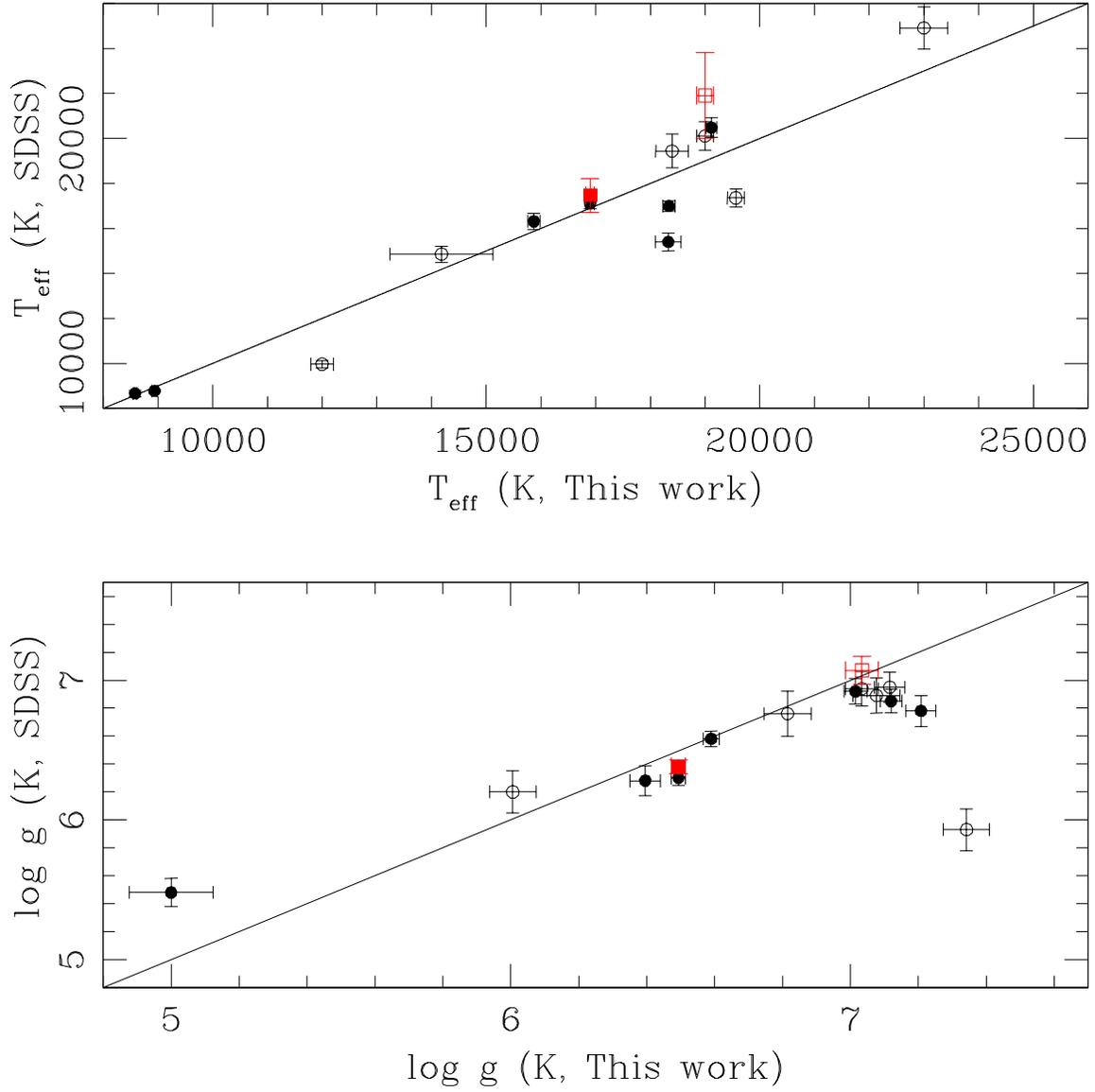}                
\caption{Eisenstein et al. (2006; circles) and Liebert et al. (2004; squares) fits to temperatures and gravities of ELM WD candidates found in the SDSS compared to our fits to the same stars. Objects with low S/N spectra (S/N $\leq$ 10 in the $g$-band) are shown as open circles and squares.}
\end{figure}

\begin{figure}
\plotone{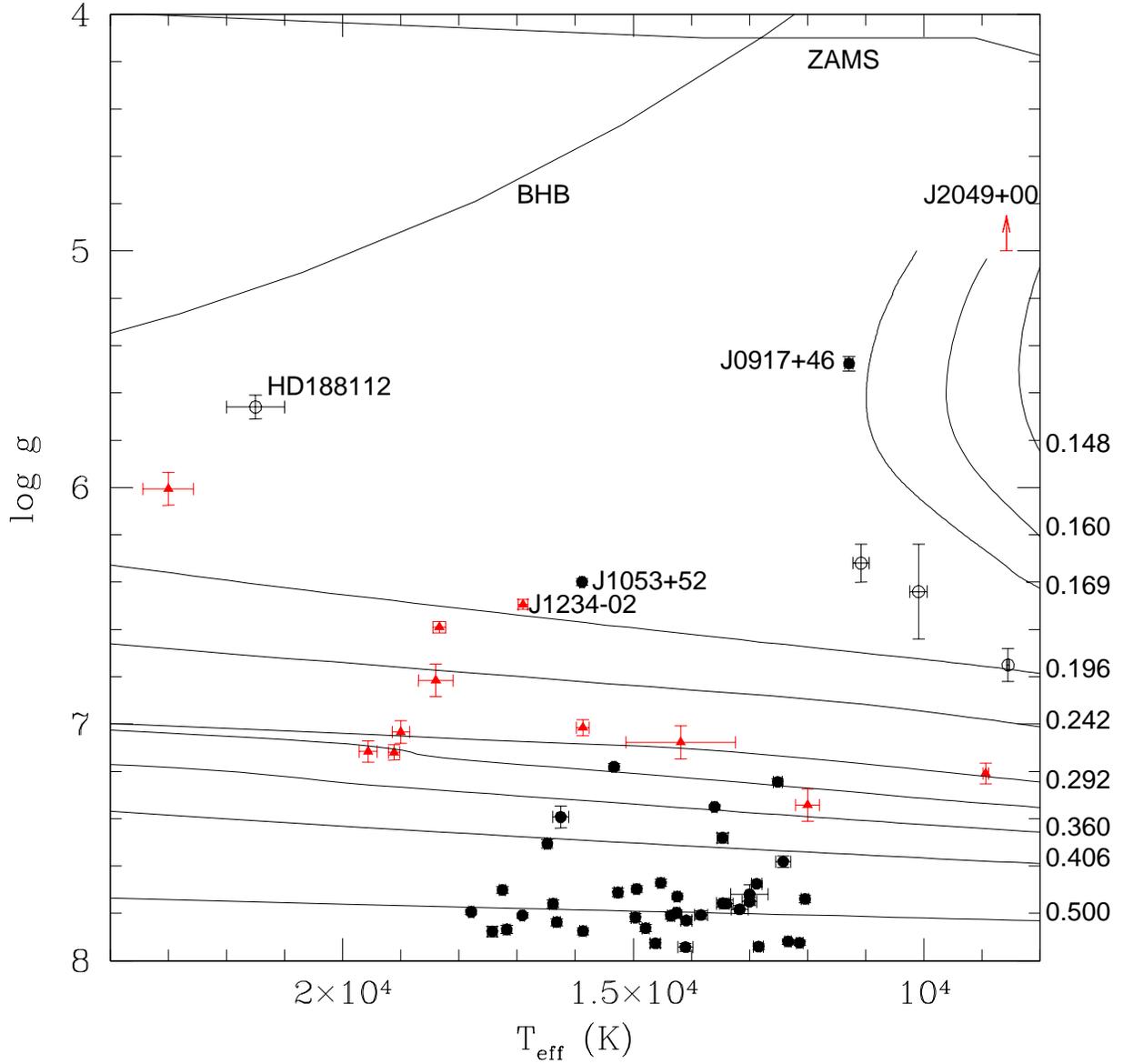}                
\caption{Our best fit solutions for the surface gravity and temperatures of the low mass WD candidates observed at the MMT
(filled circles) and the SDSS (Eisenstein et al. 2006; triangles), overlaid on tracks of constant mass from Althaus et al. (2001).
Zero-age main sequence (ZAMS) and horizontal branch star (BHB) tracks are also shown.
Spectroscopically confirmed ELM WDs found in the literature (HD 188112, LP400-22, and companions to PSR J1012+5307
and PSR J1911-5958A) are shown as open circles. }
\end{figure}


\begin{thebibliography}{}

\bibitem[Allende Prieto (2004)]{all04} Allende Prieto, C., AN, 325, 604
\bibitem[Althaus et al.(2001)]{2001MNRAS.323..471A} Althaus, L.~G., Serenelli, A.~M., \& Benvenuto, O.~G.\ 2001, \mnras, 323, 471
\bibitem[Althaus \& Benvenuto(2000)]{2000MNRAS.317..952A} Althaus, L.~G., \& Benvenuto, O.~G.\ 2000, \mnras, 317, 952 
\bibitem[Bassa et al.(2006)]{2006A&A...456..295B} Bassa, C.~G., van Kerkwijk, M.~H., Koester, D., \& Verbunt, F.\ 2006, \aap, 456, 295
\bibitem[Benvenuto \& De Vito(2005)]{2005MNRAS.362..891B} Benvenuto, O.~G., \& De Vito, M.~A.\ 2005, \mnras, 362, 891
\bibitem[Bergeron et al.(2001)]{2001ApJS..133..413B} Bergeron, P., Leggett, S.~K., \& Ruiz, M.~T.\ 2001, \apjs, 133, 413
\bibitem[Brown et al.(2006)]{2006ApJ...647..303B} Brown, W.~R., Geller, M.~J., Kenyon, S.~J., \& Kurtz, M.~J.\ 2006, \apj, 647, 303
\bibitem[Chiba \& Beers(2000)]{2000AJ....119.2843C} Chiba, M., \& Beers, T.~C.\ 2000, \aj, 119, 2843
\bibitem[Clewley et al.(2002)]{2002MNRAS.337...87C} Clewley, L., Warren, S.~J., Hewett, P.~C., Norris, J.~E., Peterson, R.~C., \& Evans, N.~W.\ 2002, \mnras, 337, 87
\bibitem[Dupuis et al.(1993)]{1993ApJS...87..345D} Dupuis, J., Fontaine, G., \& Wesemael, F.\ 1993, \apjs, 87, 345 
\bibitem[Eisenstein et al. (2006)]{eis06} Eisenstein, D. J. et al. 2006, \apjs, in press (astro-ph/0606700)
\bibitem[Ergma et al.(1998)]{1998MNRAS.300..352E} Ergma, E., Sarna, M.~J., \& Antipova, J.\ 1998, \mnras, 300, 352
\bibitem[Finley et al.(1997)]{1997ApJ...488..375F} Finley, D.~S., Koester, D., \& Basri, G.\ 1997, \apj, 488, 375
\bibitem[Gianninas et al.(2004)]{2004ApJ...617L..57G} Gianninas, A., Dufour, P., \& Bergeron, P.\ 2004, \apjl, 617, L57
\bibitem[Heber et al.(2003)]{2003A&A...411L.477H} Heber, U., Edelmann, H., Lisker, T., \& Napiwotzki, R.\ 2003, \aap, 411, L477 
\bibitem[Kawka et al.(2006)]{2006ApJ...643L.123K} Kawka, A., Vennes, S., Oswalt, T.~D., Smith, J.~A., \& Silvestri, N.~M.\ 2006, \apjl, 643, L123
\bibitem[Kilic et al.(2006)]{2006ApJ...646..474K} Kilic, M., von Hippel, T., Leggett, S.~K., \& Winget, D.~E.\ 2006, \apj, 646, 474
\bibitem[Kilic \& Redfield (2007)]{kil07} Kilic, M. \& Redfield, S. 2007, \apj, in press
\bibitem[Kleinman et al.(2004)]{2004ApJ...607..426K} Kleinman, S.~J., et al.\ 2004, \apj, 607, 426
\bibitem[Koester et al.(2005)]{2005A&A...432.1025K} Koester, D., Rollenhagen, K., Napiwotzki, R., Voss, B., Christlieb, N., Homeier, D., \& Reimers, D.\ 2005, \aap, 432, 1025
\bibitem[Liebert et al.(2005)]{2005ApJS..156...47L} Liebert, J., Bergeron, P., \& Holberg, J.~B.\ 2005, \apjs, 156, 47
\bibitem[Liebert et al.(2004)]{2004ApJ...606L.147L} Liebert, J., Bergeron, P., Eisenstein, D., Harris, H.~C., Kleinman, S.~J., Nitta, A., \& Krzesinski, J.\ 2004, \apjl, 606, L147
\bibitem[L{\"o}hmer et al.(2005)]{2005ApJ...621..388L} L{\"o}hmer, O., Lewandowski, W., Wolszczan, A., \& Wielebinski, R.\ 2005, \apj, 621, 388 
\bibitem[Marsh et al.(1995)]{1995MNRAS.275..828M} Marsh, T.~R., Dhillon, V.~S., \& Duck, S.~R.\ 1995, \mnras, 275, 828
\bibitem[Massey et al.(1988)]{1988ApJ...328..315M} Massey, P., Strobel, K., Barnes, J.~V., \& Anderson, E.\ 1988, \apj, 328, 315
\bibitem[Nelder \& Mead (1965)]{nel65} Nelder, J., \& Mead, R. 1965, Comput. J. 7, 308
\bibitem[Nice et al.(2005)]{2005ApJ...634.1242N} Nice, D.~J., Splaver, E.~M., Stairs, I.~H., L{\"o}hmer, O., Jessner, A., Kramer, M., \& Cordes, J.~M.\ 2005, \apj, 634, 1242
\bibitem[Press et al. (1986)]{pre86} Press, W. H., Flannery, B. P., Teukolsky, S. A., \& Vetterling, W. T., Numerical recipes: the art of scientific computing. Cambridge University Press, 1986
\bibitem[Rappaport et al.(1987)]{1987ApJ...322..842R} Rappaport, S., Ma, C.~P., Joss, P.~C., \& Nelson, L.~A.\ 1987, \apj, 322, 842 
\bibitem[Ritter \& Kolb(2003)]{2003A&A...404..301R} Ritter, H., \& Kolb, U.\ 2003, \aap, 404, 301
\bibitem[Schlegel et al.(1998)]{1998ApJ...500..525S} Schlegel, D.~J., Finkbeiner, D.~P., \& Davis, M.\ 1998, \apj, 500, 525 
\bibitem[Shapiro(1964)]{1964PhRvL..13..789S} Shapiro, I.~I.\ 1964, Physical Review Letters, 13, 789
\bibitem[Stella et al.(1987)]{1987ApJ...312L..17S} Stella, L., Priedhorsky, W., \& White, N.~E.\ 1987, \apjl, 312, L17
\bibitem[van Kerkwijk et al.(2000)]{2000ApJ...530L..37V} van Kerkwijk, M.~H., Bell, J.~F., Kaspi, V.~M., \& Kulkarni, S.~R.\ 2000, \apjl, 530, L37 
\bibitem[van Kerkwijk et al.(1996)]{1996ApJ...467L..89V} van Kerkwijk, M.~H., Bergeron, P., \& Kulkarni, S.~R.\ 1996, \apjl, 467, L89 
\bibitem[van Leeuwen et al. (2006)]{van06} van Leeuwen, J., Ferdman, R. D., Meyer, S., \& Stairs, I. 2006, \mnras, in press (astro-ph/0611082)
\bibitem[Zuckerman et al.(2003)]{2003ApJ...596..477Z} Zuckerman, B., Koester, D., Reid, I.~N., H\"unsch, M.\ 2003, \apj, 596, 477
\end{thebibliography}
\end{document}